\documentclass[prd,twocolumn,groupedaddress,showpacs,nofootinbib]{revtex4-1}
\usepackage{graphicx}
\usepackage{dcolumn}
\usepackage{amssymb}
\usepackage{mathrsfs}
\usepackage{amsmath}
\usepackage{epsfig}
\usepackage[dvips]{color}
\usepackage{hhline}

\begin{document}

\begin{flushright}
UMD-PP-017-23
\end{flushright}

\title{ Quark Seesaw, Dark $U(1)$ symmetry and Baryon-Dark Matter Coincidence}

\author{Pei-Hong Gu$^{1}$}
\email{peihong.gu@sjtu.edu.cn}

\author{Rabindra N. Mohapatra$^2$}
\email{rmohapat@umd.edu}

\affiliation{$^1$Department of Physics and Astronomy, Shanghai Jiao Tong
University, 800 Dongchuan Road, Shanghai 200240, China\\
$^2$Maryland Center for Fundamental Physics and Department of Physics, 
University of Maryland, College Park, Maryland 20742}

\begin{abstract} 

We attempt to understand the baryon-dark-matter coincidence problem within the quark seesaw extension of the standard model where parity invariance is used to solve the strong CP problem. The  $SU(2)_L^{}\times SU(2)_R^{}\times U(1)_{B-L}^{}$ gauge symmetry of this model is extended by a dark $U(1)_X^{}$ group plus inclusion of a heavy neutral vector-like fermion $\chi_{L,R}^{}$ charged under the dark group which plays the role of dark matter. All fermions are Dirac type in this model. Decay of heavy scalars charged under $U(1)_X^{}$ leads to simultaneous asymmetry generation of the dark matter and baryons after sphaleron effects are included. The $U(1)_X^{}$ group not only helps to stabilize the dark matter but also helps in the elimination of the symmetric part of the dark matter via $\chi-\bar{\chi}$ annihilation. For dark matter mass near the proton mass, it explains why  the baryon and dark matter abundances are of similar magnitude (the baryon-dark-matter coincidence problem). This model is testable in low threshold (sub-keV) direct dark matter search experiments.

\end{abstract}


\maketitle

\section{ Introduction}

Understanding the origin of matter in the universe and the identification of the particle physics candidate for dark matter (DM) are two major focus areas of research in particle physics today.  An important puzzle in this field whose resolution could provide key insight into these two questions is: why DM and baryon contributions to the energy budget of the universe are so close to each other or why is $\Omega_{DM}\sim 5 \Omega_B$?  This is often called the baryon-DM coincidence. A very plausible starting approach to this problem is to assume that there is a common origin of matter and DM \cite{common}. Normally this would imply that the two abundances are of similar order and if we supplement this result with the assumption that the DM particle mass is of order of the baryon mass, we arrive at the sought after resolution. The details of the mechanisms for the common origin can be different. Broadly speaking, there are two classes of models: (i) one, where the DM is a WIMP whose relic density is of thermal origin and it has interactions with the particles responsible for baryogenesis~\cite{wimpy}. This class of models are usually called "Wimpy baryogenesis" models; (ii) the second class of models where, the DM is asymmetric~\cite{nuss} and its relic density generation is similar to baryons; in this case, there may be a common origin both type asymmetries~\cite{common2}. There are constraints on both types of models: e.g. in the second class, the DM must have interactions which help to eliminate the symmetric part of the dark particle contribution to $\Omega_{DM}$ via dark particle and antiparticle annihilation. Such interactions have other cosmological implications such as the core-cusp problem, small scale structure problem etc. that arise in collisionless DM models. 

In this note we present a new model for baryon-DM coincidence puzzle which also has the additional advantage that it solves the strong CP problem using parity invariance~\cite{PCP,babumoh} within a quark seesaw framework~\cite{babumoh}. We extend the quark seesaw model by adding a dark $U(1)_X$ gauge symmetry and a vector-like fermion $\chi_{L,R}$ that has nonzero quantum numbers under this symmetry. This does not affect the strong CP solution and provides a new candidate for dark matter in the model. The new fermion $\chi$ is naturally long lived and plays the role of DM. The generation of DM and baryons comes from the same source and the symmetries of the model  allow us to relate the excesses in both baryons and DM. Using this, we find that for a DM mass $m_\chi\sim 1.33 m_p$, we get the right $\Omega_B$ and $\Omega_{DM}$. Consistency of the model requires the existence of a light vector boson with mass less than a GeV which helps in the depletion of the symmetric part of the DM as shown below and can also couple to SM fermions via kinetic mixing. The latter can eventually lead to possible direct detection of the DM in our model.

While dark $U(1)$ extensions of the standard model (SM) have been considered in the literature as a way to accommodate the DM (see for instance ~\cite{darku1}),  our model is different in many ways: we have an asymmetric DM and to have the connection between matter and DM abundance, it is essential that we have the $SU(2)_R$ group, and Dirac seesaw for neutrino masses, which together help in keeping the right handed neutrinos in equilibrium even for Dirac neutrinos and connecting the baryon excess to DM excess. Furthermore, it accommodates the Dirac neutrino mass without using excessively small Yukawa coupling (as would be the case in SM extensions) and simultaneously solve the strong CP problem.

This paper is organized as follows: in Sec. II, we present the outline of the model; in Sec. III, we discuss how the baryon asymmetry and DM asymmetry arises in our model, providing the first step towards solving the baryon-DM coincidence puzzle. In Sec. IV, we discuss the cosmology and phenomenology of the DM such as its direct detection possibility, its decay etc and in Sec.V and VI  we conclude with a brief discussion and summary of our results.

\section{The model}

We start by briefly reviewing the quark seesaw extension~\cite{QS} of the left-right symmetric model~\cite{LR}. The model is based on the gauge group $SU(3)_c\times SU(2)_L\times SU(2)_R \times U(1)_{B-L}$ gauge group. An attractive feature of the model is its simpler Higgs content, which consists of only two $SU(2)$ doublets $\phi_L^{}(1,2,1,+1)$ and $\phi_R^{}(1,1,2,+1)$, while the fermion sector contains not only the usual $SU(2)$ doublets of the left-right model i.e. $q_L^{}(3,2,1,+\frac{1}{3})$, $q_R^{}(3,1,2,+\frac{1}{3})$, $l_L^{}(1,2,1,-1)$ and $l_R^{}(1,1,2,-1)$, but also additional $SU(2)$ singlets $D_{L,R}^{}(3,1,1,-\frac{2}{3})$, $U_{L,R}^{}(3,1,1,+\frac{4}{3})$, $E_{L,R}^{}(1,1,1,-2)$ and $N_{L,R}^{}(1,1,1,0)$. Here the brackets following the fields describe the transformations of the fields under the left-right gauge group above. The $SU(2)_{L,R}^{}$ and $U(1)_{B-L}^{}$ gauge couplings are related to the standard model gauge couplings as
$g_L^{}=g$ and $1/g'^2_{}=1/g_R^2+1/g_{B-L}^2$.
The relevant Yukawa couplings and mass terms are 
\begin{eqnarray}
\label{universal}
\mathcal{L}&\supset&- (y_D^{L})_{ij}^{}\bar{q}_{Li}^{}\phi_L^{} D_{Rj}^{} -  (y_D^{R})_{ij}^{}\bar{q}_R^{}\phi_R^{}D_{Li}^{}\nonumber\\
&&-(M_D^{})_{ij}\bar{D}_{Li}^{}D_{Rj}^{}- (y_U^{L})_{ij}^{}\bar{q}_{Li}^{}\tilde{\phi}_L^{} U_{Rj}^{} \nonumber\\
&&-(y_U^{R})_{ij}^{} \bar{q}_{Ri}^{}\tilde{\phi}_R^{}U_{Lj}^{}-(M_U^{})_{ij}^{}\bar{U}_{Li}^{}U_{Rj}^{}\nonumber\\
&& - (y_E^{L})_{ij}^{}\bar{l}_{Li}^{}\phi_L^{} E_{Rj}^{} -(y_E^{R})_{ij}^{} \bar{l}_{Ri}^{}\phi_R^{}E_{Lj}^{}\nonumber\\
&&-(M_E^{})_{ij}^{}\bar{E}_{Li}^{}E_{Rj}^{}- (y_N^{L})_{ij}^{}\bar{l}_{Li}^{}\tilde{\phi}_L^{} N_{Rj}^{}\nonumber\\
&& -(y_N^{R})_{ij}^{} \bar{l}_{Ri}^{}\tilde{\phi}_R^{}N_{Lj}^{}-(M_N^{})_{ij}^{}\bar{N}_{Li}^{}N_{Rj}^{}+\textrm{H.c.}\,.
\end{eqnarray}
The model has a global lepton number symmetry. If we included Majorana mass terms for the neutral fermions $N_{L,R}^{}$, which are  allowed by the gauge symmetry,  they would explicitly violate the global lepton number by two units. In this work, we will not consider these Majorana masses of the $N_{L,R}^{}$ fermions since no lepton-number-violating processes have been observed in experiments. We will show in the next section that not only the neutrino masses  but also the charged fermion masses in this model are induced by the seesaw mechanism. The assumption of lepton number conservation has implications for DM property, as we show below.

We now extend the above universal seesaw model by a dark sector which has a dark gauge symmetry $U(1)_X^{}$ and two heavy scalars $\sigma_{1,2}^{}$, one light scalar $\xi$ and one vector-like fermion $\chi=\chi_L^{}+\chi_R^{}$. The $U(1)_X^{}$ charges for these fields are assigned as $\sigma_{1,2}^{}(+2)$, $\xi(-1)$, $\chi_{L,R}^{}(-2)$. The dark sector fields are neutral under the left-right gauge group. Besides the part for the universal seesaw, the full Lagrangian should include,
\begin{eqnarray}
\mathcal{L}&\supset& -\frac{\epsilon}{2}B_{\mu\nu}^{}X^{\mu\nu}_{}+(D_\mu^{} \sigma_{i}^{})^\dagger_{} D^\mu_{}\sigma_{i}^{} +(D_\mu^{} \xi)^\dagger_{} D^\mu_{}\xi \nonumber\\
&&+ i \bar{\chi}_{L}^{} \gamma^\mu_{} D_\mu^{} \chi_{L}^{}+ i \bar{\chi}_{R}^{} \gamma^\mu_{} D_\mu^{} \chi_{R}^{}-M_{i}^2\sigma^\dagger_{i}\sigma^{}_{i} \nonumber\\
&&- \rho_i^{}( \sigma^{}_{i}\xi^2_{}+\textrm{H.c.})- m_\chi^{}(\bar{\chi}_L^{}\chi_R^{}+\textrm{H.c.})  \nonumber\\
&&- f_{K i}^{R} \bar{N}_{LK}^{} \chi_R^{}\sigma_i^{}  - f_{K i}^{L}\bar{N}_{RK}^{} \chi_L^{}\sigma_i^{}+\textrm{H.c.}~~\textrm{with}\nonumber\\
[1mm]
&&~D_\mu^{}\sigma_{1,2}^{}= \left(\partial_\mu^{}-i 2 g_X^{}X_\mu^{}\right)\sigma_{1,2}^{}\,,\nonumber\\
&&~~~~~\!D_\mu^{}\xi= \left(\partial_\mu^{}+i g_X^{}X_\mu^{}\right)\xi\,,\nonumber\\
&&D_\mu^{}\chi_{L,R}^{}= \left(\partial_\mu^{}+i 2g_X^{}X_\mu^{}\right)\chi_{L,R}^{}\,.
\end{eqnarray}
Here $B_{\mu}^{}$ and $X_{\mu}^{}$, respectively, are the $U(1)_{B-L}^{}$ and $U(1)_X^{}$ gauge fields, while $B_{\mu\nu}^{}$ and $X_{\mu\nu}^{}$ are their strength tensors.

When the $[SU(2)_R^{}]$-doublet Higgs scalar $\phi_R^{}$ develops its vacuum expectation value (VEV) $\langle\phi_R^{0}\rangle\equiv \frac{1}{\sqrt{2}}v_R^{}$, the left-right symmetry symmetry is spontaneously broken down to the standard model electroweak symmetry. Subsequently, the $[SU(2)_L^{}]$-doublet Higgs scalar $\phi_L^{}$ acquires a VEV $\langle\phi_L^{}\rangle\equiv \frac{1}{\sqrt{2}}v_L^{}$, with $v_L^{}\simeq 246\,\textrm{GeV}$, to break the electroweak symmetry down to the electromagnetic symmetry. At this stage, the $[SU(2)]$-doublet and $[SU(2)]$-singlet fermions can have the following mass matrices, 
\begin{eqnarray}
\mathcal{L}&\supset& - \left[\begin{array}{cc}\bar{f}_L^{} & \bar{F}_L^{}\end{array}\right] M_{fF}^{}   \left[\begin{array}{c}f_R^{}  \\
[2mm]
F_R^{}\end{array}\right]  +\textrm{H.c.}~~\textrm{with}\nonumber\\
&&M_{fF}^{}= \left[\begin{array}{cc} 0  & \frac{1}{\sqrt{2}}y_F^{L}v_L^{}\\
[2mm]
 \frac{1}{\sqrt{2}}y_F^{R\dagger}v_R^{}& M_F^{}\end{array}\right]\,.
\end{eqnarray}
Here $(f,F)$ denotes $(d,D)$, $(u,U)$, $(e,E)$  and $(\nu,N)$. The $[SU(2)]$-singlet fermions $F$ are assumed heavy enough so that they can be integrated out from the above mass matrices. We hence can obtain the masses of the $[SU(2)]$-doublet fermions $f$, i.e.
\begin{eqnarray}
m_f^{} = - y_F^L \frac{v_L^{}v_R^{}}{2M_F^{}}y_F^{R\dagger}\,.
\end{eqnarray} 
Now not only the neutrino masses but also the charged fermion masses are induced by the seesaw mechanism. As a result, the Yukawa coupling parameters can have ``more natural" values compared to their values in the standard model. This is most noticeable for the Dirac neutrino mass, which in the standard model would have required $y_\nu\sim 10^{-12}$ whereas due to seesaw property, we need $y_N\sim 10^{-5.5}$ (for $v_R\simeq 2-3$ TeV).

Note that quark seesaw for the top quark, unlike the other fermions requires it to have a significant fraction of an $SU(2)$ singlet besides an $SU(2)_R^{}$ doublet since it is an $SU(2)_L^{}$ singlet. This has implications for flavor changing neutral current processes involving top decays. We also emphasize that  within the quark  seesaw framework, we can impose a discrete parity symmetry and then assume it to be softly or spontaneously broken. This leads to the Yukawa couplings in Eq. (\ref{universal}) to satisfy the relation $y_F^{L}=y_F^{R}=y_F^{}$ which helps to solve the strong CP problem without an axion~\cite{babumoh}.

Turning to the dark sector, we assume the dark scalar $\xi$ to have a VEV $\langle\xi\rangle\equiv \frac{1}{\sqrt{2}}v_\xi^{}$ to drive the spontaneous breaking of the $U(1)_X^{}$ dark symmetry. As for the other dark scalars $\sigma_{1,2}^{}$, they will pick up the tiny induced VEVs as below,
\begin{eqnarray}
\!\!\!\!\langle \sigma_i^{}\rangle\equiv \frac{1}{\sqrt{2}}v_{\sigma_i}^{}\,,~ v_{\sigma_{i}}^{}\simeq - \frac{\rho_{i}v_{\xi}^2}{\sqrt{2}\,M_{\sigma_i^{}}^2} \ll v_\xi^{} ~\textrm{for}~ M_{\sigma_i}^{} \gg v_\xi^{}\,.
\end{eqnarray}
Therefore, the heavy neutral fermions $N_{L,R}^{}$ will mix with the dark fermions $\chi_{L,R}^{}$, besides the neutrinos $\nu_{L,R}^{}$. After integrating out the $N_{L,R}^{}$ fermions, we have a $\chi-\nu$ mixing as below,  
\begin{eqnarray}
\mathcal{L}&\supset&-\theta_L^{}m_\chi^{}\bar{\nu}_L^{} \chi_R^{} -\theta_R^{}m_\chi^{}\bar{\nu}_R^{} \chi_L^{} +\textrm{H.c.}~~\textrm{with}\nonumber\\
&&\theta_L^{} =y_{N}^L\frac{v_L^{}v_\sigma^{}}{2M_N^{\dagger}}f^{R}_{}\simeq -\frac{v_\sigma^{}}{v_R^{}}\frac{m_\nu^{}}{m_\chi^{}}\frac{f_R^{}}{y_N^R}\,,\nonumber\\
&&\theta_R^{}=y_{N}^R\frac{v_R^{}v_\sigma^{}}{2M_N^{}}f^{L}_{}\simeq -\frac{v_\sigma^{}}{v_L^{}}\frac{m_\nu^{}}{m_\chi^{}}\frac{f_L^{}}{y_N^L}\,.
\end{eqnarray}
As we will show later this $\chi-\nu$ mixing have an interesting implication on the DM decay. 

Prior to gauge symmetry breaking, this model has two separate global symmetries: $U(1)_\chi$ and $U(1)_\ell$, where the latter is the usual lepton number symmetry.
After all gauge symmetries are broken and specially after $\sigma$ fields acquire VEVs, the above symmetries break but there is a remaining symmetry $U_{\ell+\chi}$. This guarantees that the neutrinos are Dirac fermions and also leads to relations between the asymmetries in the leptons and the $\chi$ fields, as we see below.

\section{Ordinary and dark matter-antimatter asymmetries}

The heavy dark scalars $\sigma_{1,2}^{}$ can decay into the neutral fermions $N_{L,R}^{}$, the dark fermions $\chi_{L,R}^{}$ as well as the light dark scalar $\xi$. We can easily calculate the decay widths at tree level, 
\begin{eqnarray}
\Gamma_i^{}&=&\Gamma\left[\sigma_i^{}\rightarrow N_{L(R)}^{}+\chi_{R(L)}^c\right] + \Gamma(\sigma_i^{}\rightarrow \xi^\ast_{}+\xi^\ast_{}) \nonumber\\
&=&\Gamma\left[\sigma_i^{\ast}\rightarrow N_{L(R)}^{c}+\chi_{R(L)}^{}\right] + \Gamma(\sigma_i^{\ast}\rightarrow \xi+\xi)\nonumber\\
&=&\frac{1}{16\pi}\left[\left(f^{L\dagger}_{}f^{L}_{}\right)_{ii}^{}+\left(f^{R\dagger}_{}f^{R}_{}\right)_{ii}^{}+2\frac{\rho_{i}^2}{M_{\sigma_i^{}}^2}\right]M_{\sigma_i^{}}^{}\,.
\end{eqnarray}
At one-loop level, we can obtain a CP asymmetry as below,
\begin{eqnarray}
\varepsilon_i^{}\!&=&\!\frac{\Gamma\!\left[\sigma_i^{}\rightarrow N_{L(R)}^{}+\chi_{R(L)}^c\right]\!-\!\Gamma\!\left[\sigma_i^{\ast}\rightarrow N_{L(R)}^{c}+\chi_{R(L)}^{}\right]}{\Gamma_i^{}}\nonumber\\
\!&=&\!\frac{\Gamma(\sigma_i^{\ast}\rightarrow \xi+\xi)- \Gamma(\sigma_i^{}\rightarrow \xi^\ast_{}+\xi^\ast_{})}{\Gamma_i^{}} \nonumber\\
\!&=&\!\frac{1}{4\pi}\frac{\textrm{Im}\left[\left(f^{L\dagger}_{}f^{L}_{}\right)_{ji}^{}+\left(f^{R\dagger}_{}f^{R}_{}\right)_{ji}^{}\right]}
{\left(f^{L\dagger}_{}f^{L}_{}\right)_{ii}^{}+\left(f^{R\dagger}_{}f^{R}_{}\right)_{ii}^{}+2\frac{\rho_{i}^2}{M_{\sigma_i^{}}^2}}\frac{\rho_i^{}\rho_j^{}}{M_{\sigma_j^{}}^2-M_{\sigma_i^{}}^2}\,.
\end{eqnarray}
Note the relative phase between the Yukawa couplings $f_{1K}^{L(R)}$ and $f_{2K}^{L(R)}$ can not be removed by any phase rotation. So, the above CP asymmetry can have a non-zero value as long as the CP is not conserved. This means we can obtain a lepton asymmetry $L_\chi^{}$ stored in the dark fermions $\chi_{L,R}^{}$. At the same time, the neutral fermions $N_{L,R}^{}$ for the seesaw can obtain an opposite lepton asymmetry $L_N^{}=-L_\chi^{}$. Through the subsequent $N_{L,R}^{}$ decays, the $[SU(2)]$-doublet leptons $l_{L,R}^{}$ then can inherit this lepton asymmetry, 
\begin{eqnarray}
\label{relation}
-L_\chi^{}=L_N^{}=L_{\textrm{SM}}^{i}+L_{\nu_R}^{i}\,.
\end{eqnarray}
Here $L_{\textrm{SM}}^{i}$ and $L_{\nu_R}^{i}$ denote the induced lepton asymmetries in the SM leptons and the right-handed neutrinos, respectively. Thanks to the $B-L$ conserving $SU(2)_{L,R}^{}$ sphaleron processes, the lepton asymmetry stored in the ordinary leptons $l_{L,R}^{}$ can be partially converted to a baryon asymmetry stored in the SM quarks \cite{krs1985,fy1986}. This connection between dark matter asymmetry and baryogenesis has similarity to the Dirac leptogenesis~\cite{lindner}.

Note that the heavy dark scalars $\sigma_{1,2}^{}$ should go out of equilibrium and then their decays can generate the desired lepton asymmetries $L_N^{}=-L_\chi^{}$. As an example, we can simply consider weak washout case, i.e.
\begin{eqnarray}
K_i^{}=\frac{\Gamma_i^{}}{H(T)}\left|_{T=M_{i}^{}}^{}\right.\ll1\,.
\end{eqnarray}
Here the Hubble constant $H(T)$ is given by
\begin{eqnarray}
\label{hubble}
H(T)=\left(\frac{8\pi^{3}_{}g_{\ast}^{}}{90}\right)^{\frac{1}{2}}_{}
\frac{T^{2}_{}}{M_{\textrm{Pl}}^{}}\,,
\end{eqnarray}
with $M_{\textrm{Pl}}^{}\simeq 1.22\times 10^{19}_{}\,\textrm{GeV}$ being the Planck mass and $g_{\ast}^{}=\mathcal{O}(200)$ being the relativistic degrees of freedom. The lepton asymmetries $L_N^{}=-L_\chi^{}$ then should be
\begin{eqnarray}
L_N^{}=-L_\chi^{}\simeq \left\{\begin{array}{lll}\frac{\varepsilon_1^{}+\varepsilon_2^{}}{g_\ast^{}}&\textrm{for} &M_{1}^{}\simeq M_{2}^{}\,,\\
[2mm]
\frac{\varepsilon_{1(2)}^{}}{g_\ast^{}}&\textrm{for} & M_{1(2)}^{}\ll M_{2(1)}^{}.\end{array}\right.
\end{eqnarray}
By taking $M_{\sigma_i}\sim 100\,\textrm{TeV}$, $M_{\sigma_2}-M_{\sigma_1}\sim 0.1\,\textrm{GeV}$, $\rho_i\sim 0.1\,\textrm{GeV}$, $f^{L,R}_{}\sim 10^{-6}$, we can obtain $K_{1,2}^{}\sim 0.1$, $\varepsilon_{1,2}^{}\sim 10^{-7}$ and then $L_N^{}=-L_\chi^{}\sim 10^{-(9-10)}_{}$.

We also wish to emphasize that the transformation of $L_N$ to $L_{\textrm{SM}}^{i}$ requires that the decay of $N$ take place in the TeV epoch of the universe and this would be possible only in the Dirac seesaw picture of the neutrino mass. Actually, we can quickly estimate the decay width $\Gamma_N^{}\sim y^2_N M_N^{}\sim 10^{-9}_{}\,\textrm{GeV}$ for $y^{}_{N}\sim 10^{-6}$ and $M_N^{}\sim 10\,\textrm{TeV}$. This decay width is much bigger than the Hubble constant at the temperature around $T\sim 100\,\textrm{GeV}$. In the absence of Dirac seesaw, the Yukawa couplings of the right-handed neutrinos to the SM leptons is $\sim 10^{-12}$ which allows the $N$ to decay very late in the universe by which time sphalerons have gone out of equilibrium and the $L_N$  asymmetry cannot become $B$ asymmetry.

We can calculate the final baryon asymmetry converted by an initial lepton asymmetry. For this purpose, we denote $\mu_{q,d,u,l,e,\nu,\phi}^{}$ for the chemical potentials of the SM fermions $q_{L}^{}(3,2,+1/6)$, $d_{R}^{}(3,1,-1/3)$, $u_{R}^{}(3,1,+2/3)$, $l_L^{}(1,2,-1/2)$, $e_R^{}(1,1,-1)$, the right-handed neutrinos $\nu_R^{}(1,0,0)$ and the SM Higgs scalar $\varphi(1,2,+1/2)$. Here the brackets following the fields give the $SU(3)_c^{}\times SU(2)_L^{}\times U(1)^{}_{Y}$ quantum numbers. At the electroweak scale, the SM Yukawa interactions yield \cite{ht1990},
\begin{eqnarray}
\label{smych}
&&-\mu_{q}^{}+\mu_{\varphi}^{}+\mu_{d}^{}=0\,,~~
 -\mu_{q}^{}-\mu_{\varphi}^{}+\mu_{u}^{}=0\,,\nonumber\\
 &&
 -\mu_{l}^{}+\mu_{\varphi}^{}+\mu_{e}^{}=0\,,
\end{eqnarray}
the $SU(2)_L^{}$ sphalerons constrain \cite{ht1990},
\begin{eqnarray}
\label{sphch}
3\mu_{q}^{}+\mu_{l}^{}&=&0\,,
\end{eqnarray}
while the vanishing hypercharge in the universe require \cite{ht1990},
\begin{eqnarray}
\label{hyperc}
3\left(\mu_{q}^{}-\mu_{d}^{}+2\mu_{u}^{}-\mu_{l}^{}-\mu_{e}^{}\right)+2\mu_{\varphi}^{} =0\,.
\end{eqnarray}
At this stage, the right-handed charged current interactions,
\begin{eqnarray}
\label{rcurrent}
\mathcal{L}\supset \left(\frac{m_{W_L}^2}{m_{W_R}^2}\right)\frac{G_F^{}}{\sqrt{2}} \bar{u}_R^{}\gamma^\mu_{}d_R^{} \bar{e}_R^{}\gamma_\mu^{}\nu_R^{}
+\textrm{H.c.}\,,
\end{eqnarray}
can be in equilibrium, depending on the left-right symmetry breaking scale. In this case, we further have
\begin{eqnarray}
\label{rightnu}
-\mu_{u}^{}+\mu_{d}^{}-\mu_e^{}+\mu_\nu^{}&=&0\,.
\end{eqnarray} 
This is where the $SU(2)_R$ interactions play an important role. If they did not exist, $L_{\nu_R}$ would essentially remain as an unknown parameter hindering a direct connection between $L_N$ and $B$ asymmetry.
Note that constraints on chemical potentials imposed by QCD sphalerons being in equilibrium~\cite{xinmin} is automatically satisfied in our case, using the first two equations in Eq. (\ref{smych}).

All chemical potentials can be expressed in terms of a single chemical potential. For example, we read
\begin{eqnarray}
&&\mu_q^{}=-\frac{1}{3}\mu_l^{}\,,~~\mu_d^{}=-\frac{19}{21}\mu_l^{}\,,~\mu_u^{}=\frac{5}{21}\mu_l^{}\,,~~\mu_e^{}=\frac{3}{7}\mu_l^{}\,,\nonumber\\
&&\mu_\nu^{}=\frac{11}{7}\mu_l^{}\,,~~\mu_\varphi^{}=\frac{4}{7}\mu_l^{}\,.
\end{eqnarray} 
The corresponding baryon and lepton asymmetries then should be 
\begin{eqnarray}
B&=&3\left(2\mu_q^{}+\mu_u^{}+\mu_d^{}\right)=-4\mu_l^{}\,,\nonumber\\
L_{\textrm{SM}}^{}&=&3\left(2\mu_l^{}+\mu_e^{}\right)=\frac{51}{7}\mu_l^{}\,,\nonumber\\
L_{\nu_R}^{}&=&3\mu_\nu^{}=\frac{33}{7}\mu_l^{}\,.
\end{eqnarray}

If the right-handed charged current interactions (\ref{rcurrent}) have not decoupled before the $SU(2)_L^{}$ sphaleron processes stop working, as is the case for TeV scale $W_R$ models, we can easily read the final baryon asymmetry from the initial lepton asymmetries,
\begin{eqnarray}
\label{result1}
B=\frac{1}{4}\left[B^i_{}-\left(L_{\textrm{SM}}^{i}+L_{\nu_R}^{i}\right)\right]= \frac{1}{4}L_\chi^{}\,,
\end{eqnarray} 
by adopting the relation (\ref{relation}) and assuming the initial baryon asymmetry to be zero, i.e. $B^i_{}=0$.

\section{Dark matter phenomenology}

The cosmological observations have precisely measured the energy densities of the baryonic and DM in the present universe, i.e. $\Omega_{b}^{}h^2_{}=0.02226\pm 0.00023$, $\Omega_{\chi}^{}h^2_{}=0.1186\pm 0.0020 $\cite{patrignani2016}.
If the DM relic density is a dark matter-antimatter asymmetry, the dark fermion should have a special mass to match the observations, i.e.
\begin{eqnarray}
m_p^{} B : m_\chi^{} L_\chi^{}=\Omega_{b}^{}h^2_{}:\Omega_{\chi}^{}h^2_{}\,.
\end{eqnarray}
The DM mass thus can be predicted by 
\begin{eqnarray}
m_\chi^{} = \frac{1}{4}\frac{\Omega_{\chi}^{}h^2_{}}{\Omega_{b}^{}h^2_{}} m_p^{}=1.332\,m_p^{}\left(\frac{\Omega_{\chi}^{}h^2_{}/0.1186}{\Omega_{b}^{}h^2_{}/0.02226}\right)\,.
\end{eqnarray}

\subsection{Annihilation of the symmetric part of dark matter }

It is well known that an asymmetric DM scenario must require a fast dark matter-antimatter annihilation to highly suppress the thermally produced DM relic density. This can be easily achieved in the present model where the dark fermion $\chi$ couples to the dark photon $X_\mu^{}$ with a mass $m_X^{2}=g_X^{2}(v_\xi^2+4v_{\sigma_{1,2}^{}}^2)\simeq g_X^{2}v_\xi^2$. Specifically, a dark fermion pair can annihilate into two dark photons by the exchange of a dark fermion as well as into one dark photon and one dark Higgs boson, which is mainly form the light dark scalar $\xi$, through the exchange of a dark photon if the dark photon and the dark Higgs are lighter than the dark fermion, i.e.
\begin{eqnarray}
\langle\sigma_{\textrm{A}}^{} v_{\textrm{rel}}^{}\rangle\simeq\frac{g_X^4}{\pi}\frac{1}{m_\chi^2}= 9.3\times 10^{3}_{}\,\textrm{pb} \left(\frac{g_X^{}}{0.1}\right)^4_{}\,.
\end{eqnarray}
Clearly, the dark matter-antimatter annihilation can be very fast unless the $U(1)_X^{}$ gauge coupling is extremely small and it implies $g_X\geq 0.01$ for $m_X\sim $ GeV. This helps to suppress the symmetric part of the DM leaving only the asymmetric part.

Furthermore, the dark gauge boson can mediate a DM self-interaction. The cross section is given by
\begin{eqnarray}
\frac{\sigma_{\textrm{self}}^{}}{m_\chi^{}}&=&\frac{8g_X^4}{\pi}\frac{m_\chi^{}}{m_{X}^4}= \frac{8}{\pi}\frac{m_\chi}{v_\xi^4}\nonumber\\
&=&8.7\times 10^{-50}_{}\,\textrm{cm}^2_{}\cdot \textrm{sec}\left(\frac{1\,\textrm{GeV}}{v_\xi^{}}\right)^4_{}\,.
\end{eqnarray}
which is safely below the current limits from bullet cluster on the cross section which is of order $\sigma_{\rm self}\leq 10^{-25}$ cm$^2$ for a one GeV DM \cite{wgd2017}.

\subsection{Direct detection}

In the absence of the kinetic mixing between the $U(1)_{B-L}^{}$ and $U(1)_X^{}$ gauge fields (i.e. when $\epsilon=0$), there is no interaction between the DM field and the SM fields. Once the $U(1)_{B-L}^{}$ and $U(1)_X^{}$ kinetic mixing is turned on, the dark gauge field $X_\mu^{}$ can couple to the ordinary charged fermions $d,u,e$ in addition to the dark fermion $\chi$, i.e.
\begin{eqnarray}
\mathcal{L}&\supset& -\frac{\epsilon e}{\sqrt{\cos 2\theta_W^{}}} \left(-\frac{1}{3}\bar{d}\gamma^\mu_{}d+\frac{2}{3}\bar{u}\gamma^\mu_{}u -\bar{e}\gamma^\mu_{}e \right)X_\mu^{}\nonumber\\
&&-2g_X^{}\bar{\chi}\gamma^\mu_{}\chi X_\mu^{}\,.
\end{eqnarray}
The dark photon thus may be found at colliders. The dark photon can also mediate a spin-independent scattering of the dark fermion off the protons in nuclei, 
\begin{eqnarray}
\sigma_{\textrm{SI}}^{} &=&\frac{16\alpha \epsilon^2_{}g_X^2}{\cos2\theta_W^{}}\frac{\left[m_p^{}m_\chi^{}/(m_p^{}+m_\chi^{}\right]^2_{}}{m_X^4 }\nonumber\\
&=&1.2\times 10^{-39}_{}\,\textrm{cm}^2_{} \nonumber\\
&&
\times \left(\frac{\epsilon}{10^{-6}_{}}\right)^2_{}\left(\frac{100\,\textrm{MeV}}{m_X^{}}\right)^2_{}\left(\frac{1\,\textrm{GeV}}{v_\xi^{}}\right)^2_{}\,.
\end{eqnarray}

Such low mass of the dark matter, as predicted by our model is not accessible to any of the currently running experiments  in the DM-proton scattering mode. CRESST-II experiment presents results for dark matter mass slightly above one GeV and puts a bound on the spin independent cross section at $10^{-36}$ cm$^2$~\cite{CRESST} and planned  CRESST-III~\cite{CRESST} experiment will have the ability to improve the sensitivity further. The superCDMSSNOLAB  experiment, which is a proposal for a second generation experiment~\cite{SCDMS} is also planned to cover this low mass range in the DM-proton scattering mode. We hasten to point out, our predictions for DM-proton scattering depends on the unknown photon-X-boson mixing parameter $\epsilon$ and any improvement of the low mass DM search experiment will simply constrain this parameter.
 
\subsection{Dark matter decay}

The mixing between DM fermion $\chi$ and neutrinos $\nu_{L}^{}$ will lead to a tree-level decay of DM into a neutrino and a dark gauge photon, X. Fortunately, the dark fermion can have a very long lifetime because of the extremely tiny $\chi-\nu$ mixing parameter $\theta_{L,R}$. The dominant contribution comes from $\theta_{R}$. In order to estimate this decay time, we choose, $M_{\sigma_i}\sim 100\,\textrm{TeV}$, $M_F\sim 10\,\textrm{TeV}$, $\rho_i\sim 0.1\,\textrm{GeV}$, $f\sim 10^{-6}$, which leads to $v_\sigma\sim 10^{-12}\,\textrm{GeV}$ and the resulting $\theta_R\sim  10^{-24}$, 
\begin{eqnarray}
\frac{1}{\tau_\chi^{}}&=&\frac{g_X^2}{\pi}\left(\theta_L^2+\theta_R^2\right) m_X^{}=\frac{1}{1.6\times 10^{27}_{}\,\textrm{sec}}\left(\frac{\theta_L^2+\theta_R^2}{10^{-48}_{}}\right)\,.\nonumber\\
&&
\end{eqnarray}  
The dark photons from the DM decays then can decay into the electron-positron pairs and probably can also decay into two mesons if the dark photon is heavy enough. 
The effect of such late time dark matter decays have been studied in ~\cite{DMDECAY}, where a lower limit on the lifetime of $\gtrsim 2\times 10^{25}$ sec. has been obtained based on current Planck, WMAP9, SPT and ACT, as well as Lyman-$\alpha$ measurements. Our model clearly satisfies this bound.

\section{Other implications}

We now make  a few comments on other implications of our model:
\begin{itemize}

\item The right-handed components of the Dirac neutrinos will contribute to the effective number of additional light neutrinos, $\Delta N_{eff}$, an issue that has been discussed in the literature~\cite{barger}. The key point is the decoupling temperature of the right handed neutrinos which couple to the SM fermions via the $W_R$ and $Z'$ mediated interactions. This implies a lower limit on the $W_R$ and $Z'$ boson in the range of a few TeVs~\cite{barger}. 

\item When the kinetic mixing parameter is chosen to be non-zero, the extra light dark gauge boson will have interactions with the muon and contribute to the $g_\mu-2$ of the muon. However, in our model, we have chosen the $\epsilon\simeq 10^{-6}$, its contribution is much smaller than the current uncertainties in its measured value.

\item For nonzero $\epsilon$, the dark gauge boson will couple to electrons and neutrinos in a supernova and could be produced if its mass is $\leq 100$ MeV. We keep the $m_X\geq 100$ MeV so that we do not face this constraint.
For other limits on the kinetic mixing parameter for a given dark photon mass, see~ \cite{best2009}.

\item Finally, we note that quark seesaw models have many interesting phenomenological  signatures~\cite{pheno} including new heavy quarks and leptons with TeV mass, their effects on flavor changing decays of the top quark etc that have been extensively discussed in the literature. Typically the effect is largest in top quark decays e.g. $t\to c,u+g, c,u+\gamma, t\to c,u+Z$ ($g$ stands for gluons). Such decays have been searched for in collider experiments~\cite{tf}.

\end{itemize}

\vskip0.4in

\section{Comments and conclusion}

In summary, we have presented a simple extension of the $SU(3)_c\times SU(2)_L\times SU(2)_R \times U(1)_{B-L}$ left-right symmetric models for quark seesaw by including a dark $U(1)_X^{}$ and a Dirac dark fermion that relates the abundance of DM and baryons and thereby explains the baryon-DM coincidence problem. The complete model has a global $U(1)_{\ell+\chi}$ conservation, so that asymmetry in the DM is related by this symmetry to the asymmetry in the lepton sector which via both $SU(2)_{L,R}$ sphalerons gets converted to baryon asymmetry. The generation of matter and DM is caused by the decay of two heavy scalars $\sigma_{1,2}^{}$ that carry the $U(1)_\chi$ quantum number. We choose their masses to be in the $10\,\textrm{TeV}$ range but they could be superheavy in which case, they can play the role of inflaton. The model has the additional advantage over just extending the SM by a dark $U(1)$ that it does not require ultra-low Yukawa couplings to understand neutrino masses and it solves the strong CP problem. For certain ranges of the $U(1)_{B-L}^{}\times U(1)_X^{}$ kinetic mixing, the DM can give a signal in the direct detection experiments.

\textbf{Acknowledgement:} The work of P.H.G. was supported by the National Natural Science Foundation of China under Grant No. 11675100, the Recruitment Program for Young Professionals under Grant No. 15Z127060004, the Shanghai Jiao Tong University under Grant No. WF220407201, the Shanghai Laboratory for Particle Physics and Cosmology under Grant No. 11DZ2260700 and the Key Laboratory for Particle Physics, Astrophysics and Cosmology, Ministry of Education. The work of R.N.M. was supported by the US National Science Foundation under Grant No. PHY1620074.

\end{document}